\documentstyle[twocolumn,aps,epsfig]{revtex}
\begin{document}

\title{Stopped Muon Decay Lifetime Shifts due to Condensed Matter}

\author{A. Widom[1], Y. Srivastava[1,2] and J. Swain[1]}
\address{1. Physics Department, Northeastern University, Boston MA 
02115 U.S.A.}
 
\address{2. Physics Department \& INFN, University of Perugia, Perugia, 
Italy}
 
\maketitle

\begin{abstract}                                                           
Up to second order in $\alpha =(e^2/\hbar c)$,
vacuum electro-magnetic corrections to weak interaction induced   
charged particle lifetimes have been previously studied. In the 
laboratory, stopped muon lifetimes are measured in a condensed 
matter medium whose radiation impedance differs from that of the 
vacuum. The resulting condensed matter corrections to first order 
in $\alpha $ dominate those vacuum radiative corrections (two photon 
loops) which are second order in $\alpha $.
\end{abstract}  


For unstable charged particles, such as the muon, electro-magnetic 
corrections to the lifetime are essential for precise determinations
of weak interaction coupling strengths. For example, tests of 
universality for lepton couplings to the weak current are meaningful 
only after radiative corrections have been applied. The needed 
calculations have a long history \cite{1,2,3}, 
and investigations into the limit $(m_e/M_\mu)\to 0^+ $ were quite 
fruitful. These investigations eventually led to wonderful insights 
into the nature of mass singularities, e.g. the Kinoshita-Lee-Nauenberg 
theorem \cite{4,5}, and their applications to jet definitions\cite{6,7}. 
More recently, computations of two-loop photon corrections to second 
order in $\alpha =(e^2/\hbar c)$ have also been made\cite{8}.

All of the above calculations are applicable to muons which decay in 
the vacuum. In {\em real experiments}, the muons are stopped in 
condensed matter before they decay. Curiously, and even though 
condensed matter environment related life time shifts of decaying 
nuclei have been experimentally confirmed \cite{9}, we have been 
unable to find similar studies for weak charged particle decays. Such an 
effect must exist since the radiation spectrum (emitted during a charged 
particle decay) depends on the nature of material environment.  
In what follows, we shall exhibit the radiative corrections to the 
weak decay rate for muons both in condensed matter and in the vacuum. 
We conclude, for {\em laboratory experiments}, that condensed matter 
effects to first order in $\alpha $ dominate those vacuum effects which 
are of second order in $\alpha $.

To lowest order in electro-weak theory, the Fermi transition 
rate $\Gamma_F$ for the muon decay $\mu^-\to e^- +\bar{\nu}_e+\nu_\mu$ 
yields 
\begin{equation}
\Gamma_F=\left({M_\mu c^2\over 192\pi ^3\hbar}\right)
\left({G_FM_\mu^2\over \hbar c}\right)^2 
\left\{1-8\left({m_e\over M_\mu }\right)^2\right\}.
\end{equation}
The first Feynman diagram of FIG.1 corresponds to Eq.(1) 
The radiative corrections to lowest order in $\alpha $ are described 
by the five remaining Feynman diagrams.

\begin{figure}[htbp]
\begin{center}
\mbox{\epsfig{file=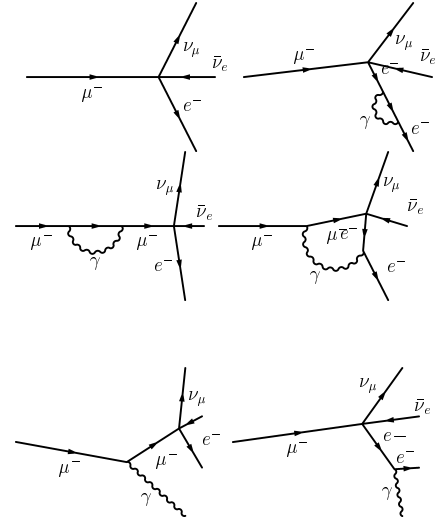,height=100mm}}
\caption{Shown are (i) the Fermi interaction diagram for muon decay as in 
Eq.(1), (ii) the electron wave function renormalization to first order 
in $\alpha $, (iii) the muon wave function renormalization to first order 
in $\alpha $, and (iv) the vertex renormalization to first order in 
$\alpha $. In these diagrams, photons are virtual. Shown in (v) and (vi) 
are the ``real'' Bremsstrahlung photon emission diagrams, respectively, 
from the muon and electron.}
\label{fig1}
\end{center}
\end{figure}
 
One computes from the amplitudes in FIG.1, the radiative corrections to 
$\Gamma_F$ to lowest order in $\alpha $. The final answer for the total 
decay rate is   
\begin{equation}
\Gamma_{tot}= \Gamma_F 
\left\{ 
1 - \left({\alpha \over 2\pi }\right)
\left(\pi^2 - \left({25 \over 4}\right) + \eta \right)+\ ...\ 
\right\},
\end{equation} 
where the parameter $\eta $ will be used to describe the condensed 
matter effects. 

For the vacuum case with $\eta =0$, Eq.(2) is well known. The vacuum 
radiative corrections slightly suppress the muon decay rate. The 
physical reason is that the virtual photon diagrams (when interfering 
with the zeroth order amplitude) subtract from the transition rate. 
The Bremsstrahlung emission of real photons adds to the decay rate 
by introducing a new photon channel. However, the elastic subtraction 
wins out and determines the overall sign of the effect.

The rules of quantum electrodynamics for the diagrams in FIG.1 require 
the propagator (most often) written in the Feynman gauge. For the vacuum  
\begin{equation}
D^{vac}_{\mu \nu }({\bf k},\omega )=\left(
{4\pi \eta_{\mu \nu} \over |{\bf k}|^2-(\omega /c)^2+i0^+}
\right).
\end{equation}
In condensed matter, wherein resides the stopped muon, the 
properties of the medium will be described by a dielectric 
response function analytic in the upper-half complex frequency 
$\zeta $ plane; i.e.   
\begin{equation}
\varepsilon (\zeta )=1+{2\over \pi}
\int_0^\infty{\omega \Im m\ \varepsilon (\omega +i0^+)d\omega 
\over \omega^2 -\zeta ^2}\ ,
\end{equation} 
where $\Im m\ \zeta >0$. In the condensed matter media, the 
Feynman gauge propagator $D_{\mu \nu }({\bf k},\omega )$ is 
still diagonal in the indices $(\mu \nu)$. However, the diagonal 
elements are given by \cite{10} 
\begin{equation}
D_{00}({\bf k},\omega )=
-\left(
{\big(4\pi /\varepsilon(|\omega |+i0^+)\big) \over 
|{\bf k}|^2-\varepsilon(|\omega |+i0^+)(\omega /c)^2+i0^+}
\right)
\end{equation}
and 
\begin{equation}
D_{ij}({\bf k},\omega )=\left(
{4\pi \delta_{ij} \over 
|{\bf k}|^2-\varepsilon(|\omega |+i0^+)(\omega /c)^2+i0^+}
\right)
\end{equation}

Eqs.(5) and (6) are required for the photon propagator insertions 
in the Feynman diagrams of FIG.1. For the Bremsstrahlung emission 
diagrams, the out going photon wave function must be renormalized 
in the following manner: (i) Define the condensed matter 
polarization part $\Pi^{\lambda \sigma }({\bf k},\omega )$ via 
\begin{equation}
D_{\mu \nu}=D^{vac}_{\mu \nu }
+D^{vac}_{\mu \lambda }\Pi^{\lambda \sigma }
D^{vac}_{\sigma \nu }.
\end{equation}
(ii) If $a_\mu ({\bf k},\omega )$ denotes the outgoing photon wave 
function in the vacuum, and if $A_\mu ({\bf k},\omega )$ denotes 
the outgoing photon wave in the condensed media, then 
\begin{equation}
A_\mu =a_\mu +D^{vac}_{\mu \sigma }\Pi^{\sigma \nu }a_\nu .
\end{equation}  
(iii) Finally, in the high frequency limit 
\begin{equation}
\varepsilon (\zeta )\to 1-\left({\omega _p^2\over \zeta ^2}\right) 
\ \ \ {\rm as}\ \ |\zeta |\to \infty 
\end{equation}
where the ``plasma frequency'' is determined by the number of electrons 
per unit volume $n_e$ via  
\begin{equation}
\omega_p^2=\left({4\pi n_e e^2\over m_e}\right).
\end{equation}
Equivalently we have the sum rule 
\begin{equation}
{2\over \pi}\int_0^\infty \omega 
\Im m\ \varepsilon (\omega +i0^+)d\omega =\omega_p^2,
\end{equation}
so that $\hbar \omega_p$ sets a photon energy scale beyond which 
the condensed matter radiation impedance is the same as the vacuum 
radiation impedance. In practice, condensed matter effects are 
important only for photons of energy less than (say) $5$ KeV. 

For a stopped muon decay, the radiation given off by the product  
electron is substantial. To zeroth oder, the distribution of the 
product electron energy $E$ is given by 
\begin{equation}
d\Gamma_F(E)=\Gamma_F dP(E), \ \ \ (0<E<W),
\end{equation}
where 
\begin{equation}
{{dP(E)}\over{dE}} = 6({{E^2}\over{W^3}}) - 4({{E^3}\over{W^4}}),
\end{equation}
and the electron energy cut-off is $W=(M_\mu c^2/2)$. If ${\bf v}$ denotes 
the electron velocity, 
\begin{equation}
E={m_ec^2\over \sqrt{1-({|\bf v}|/c)^2}},
\end{equation}
and $dN(\omega ,E)$ denotes the distribution 
of Bremsstrahlung photons in a bandwidth $d\omega $, then for the vacuum 
\begin{equation}
dN^{vac}(\omega ,E)=\beta^{vac} (E)
\left({d\omega \over \omega}\right)
\end{equation} 
where 
\begin{equation}
\beta^{vac} (E)=
\Big({\alpha \over \pi}\Big)
\Big\{
\Big({c\over |{\bf v}|}\Big)
\ln\Big({c+|{\bf v}|\over c-|{\bf v}|}\Big)-2
\Big\}.
\end{equation}
The corresponding distribution of radiated photons in the
condensed matter environment
\begin{equation}
dN(\omega ,E)=\beta  (\omega ,E)
\left({d\omega \over \omega}\right)
\end{equation} 
has been discussed elsewhere \cite{11}. The final result is given in terms 
of 
\begin{equation}
z(\omega )=\Big({c\over |{\bf v}|\sqrt{\varepsilon(\omega +i0^+ )}}\Big)
\end{equation}
as  
\begin{equation}
\beta (\omega ,E)=\left({\alpha |{\bf v}|\over c\pi}\right)
\left|{(\Re e\ z) \big(\Im m \ {\cal G}(z)\big)\over (\Im m\ z)}\right|,
\end{equation}
where 
\begin{equation}
{\cal G}(z)=\Big({z^2-1\over 2}\Big)\ln\Big({z+1\over z-1}\Big)-z.
\end{equation}
Using Eqs.(2), (5), (6), (13), (18), (19) and (20), and employing the 
condition that  $\hbar \omega <<W$ in the integral regime in which the 
material and vacuum values $\beta $ differ appreciably, one finds 
\begin{equation}
\eta=\left({2\pi  \over \alpha}\right)
\left({dP(E)\over dE}\right)_{E=W}\Delta E_{rad} 
\end{equation}
where
\begin{equation}
\Delta E_{rad}=\hbar \int_0^\infty 
\left(\beta (\omega ,E=W)-\beta^{vac} (E=W)\right)d\omega . 
\end{equation}
Our central results follow from Eqs.(2), (13), (22) and (23).
The total difference in the mean radiated energy 
$\Delta E_{rad}=E_{rad}-E^{vac}_{rad}$ between the material and the vacuum, 
at the maximum electron energy $W\approx(M_\mu c^2/2)$ determines 
the material radiation renormalization parameter 
\begin{equation}
\eta =\left({8\pi \over \alpha }\right)
\left({\big(E_{rad}-E^{vac}_{rad}\big)\over M_\mu c^2}\right)
\approx \left({\Delta E_{rad}\over 30.78\ KeV}\right).
\end{equation}

At this point it is important to distinguish three regimes for the motion 
of the charge. (i) When the muon is stopped, the velocity of the charge 
is zero. (ii) When the muon decays, there is a rapid acceleration of the 
charge to a final electron velocity ${\bf v}$. The acceleration produces 
a pulse of radiation with a mean photon distribution 
$\omega dN(\omega ,E)/d\omega =\beta (\omega ,E)$. These two 
regimes are present in the vacuum. (iii) The third (and final) 
regime is when the electron slowly decelerates with an energy loss 
as it leaves a track through the condensed matter medium. This last 
process is usually described by a retardation force \cite{12} 
\begin{equation} 
F=-\left({dE\over dx}\right)
\end{equation}
which has no counterpart in the vacuum. 

For example, consider a regime in which the material is almost 
transparent; i.e. 
\begin{equation}
\sqrt{\varepsilon (\omega +i0^+)}=n(\omega )e^{i\phi (\omega )}, 
\ \ \ \ \ \ |\phi (\omega )|<<1.
\end{equation}
where $n$ is the index of refraction and $\tan (2\phi) $ is the loss 
tangent of the dielectric response $\varepsilon (\zeta )$. In such a regime, 
and for 
\begin{equation}
|{\bf v}|>\left({c\over n}\right) 
\ \ \ \ {\rm (Cerenkov)},
\end{equation}
the Cerenkov radiation retardation force has the well known frequency 
distribution \cite{13} 
\begin{equation}
dF_C=\left({e^2\over c^2}\right)
\left(1-\left({c\over |{\bf v}|n}\right)^2\right)\omega d\omega .
\end{equation}
This corresponds to the number of Cerenkov photons emitted in a bandwidth 
$d\omega $ and in a path length $dx $ given by 
\begin{equation}
\left({d^2N_C\over d\omega dx}\right)=\left({\alpha \over c}\right)
\left(1-\left({c\over |{\bf v}|n}\right)^2\right) .
\end{equation}

Eq.(28) describes the emission of Cerenkov radiation during the 
third regime in which the electron leaves a track. The 
emission of Cerenkov radiation in the second regime (in which the 
charge rapidly accelerates from zero velocity to ${\bf v}$) is governed by 
Eqs.(18)-(20) and (25),   
\begin{equation} 
\beta_C=\left({\alpha |{\bf v}| \over 2c\tan\phi }\right)
\left(1-\left({c\over |{\bf v}|n}\right)^2\right),\ \ \ |\phi |<<1.
\end{equation}
Note the loss angle term $tan \phi $ is in the denominator of Eq.(29). 
If the material in a frequency range of interest were really 
transparent, then $\phi \to 0$ would imply the divergence  
$\beta_C=\omega (dN_C/d\omega )\to \infty $. As previously 
discussed\cite{11}, the Cerenkov contribution to $\beta $ thereby  depends 
sensitively on the attenuation of the radiated waves. In some continuous 
media detector systems, e.g. optical fibres with a very small 
electromagnetic attenuation, there should be a very large flash 
of Cerenkov radiation. 

The physical picture may be described as follows: Consider a modern jet 
plane leaving an airport. The plane starts out at rest, rapidly 
accelerates upon ``take-off'' and continues to accelerate until sound speed 
is approached. When the plane then accelerates right through the 
sound speed barrier, there follows a loud ``thunder clap'' or sound wave 
``explosion''. (A considerate airline will have the pilot avoid breaking 
the sound speed barrier while flying over a city since the resulting 
{\em sound wave explosion} would be quite unpleasant for the the city 
inhabitants to hear). After the sound speed barrier is broken, the plane 
still sends out some further, but much more mild, sound waves along the 
``Mach-Cerenkov cone''. The sounds are much more mild when flying at a 
uniform velocity than while accelerating through the sound speed barrier. 
With the sound wave analogy kept firmly in mind, let us return to the case 
of muon decay.

The muon starts off at rest and decays into an electron (plus, of course, 
uncharged objects). The electron quickly accelerates (so to speak) to a 
final velocity ${\bf v}$ which may break through the light speed 
barrier; i.e. $|{\bf v}|>(c/n)$ for some material bandwidths. When the 
light speed barrier is broken, there is an large flash of Cerenkov 
electromagnetic radiation with a photon distribution 
$(dN_C/N_C)=\beta_C(d\omega /\omega)$ according to Eq.(29). After the light 
speed barrier is broken, the electron (moving at a roughly uniform velocity) 
still sends out a further (but much more mild) Cerenkov electromagnetic 
signal in accordance with Eqs.(27) and (28). 

The crucial point is the following: If the Cerenkov flash of radiation in the 
material is sufficiently large, then the electromagnetic renormalization 
of the muon decay rate {\em due to the material} will also be quite large.
The exact numbers depend on the material properties 
$\varepsilon (\omega +i0^+)$. 

For typical plastic coated glass optical fibres, we expect, in a 
frequency bandwidth $\hbar \Delta \omega \sim 0.2\ eV$, an index 
of refraction $n\sim 1.3$. Most importantly, there will be a small 
light wave attenuation (partly from Rayleigh scattering and partly from 
infrared electronic transitions) with a 
loss angle $\phi \sim 10^{-8}$\cite{14,15}. Using these estimates,  
we find for the Cerenkov contribution to Eq.(2) $\eta_C\sim 0.9$ 
which is substantial. For muon decay in metals, the effect is 
much reduced in magnitude. We note (in passing) that metals 
suppress radiation which may change the sign of $\eta $ to 
a negative value. 

While we were led to the notion of a Cerenkov flash in optical fibres 
via the consideration of the Feynman diagrams in FIG.1, it would 
appear to us to be of scientific interest to measure such  
optical ``bolts of lightning'' in situ per se. By contrast, when a 
charged pion at rest decays in a transparent medium, the Cerenkov 
flash will be very small or virtually zero. In pion decay the muon 
emerges at a velocity which can hardly beat that of light in the material. 
When the muon decays, the flash caused by the emerging electron (moving 
almost at vacuum light speed) should be substantial and measurable. 
These Cerenkov effects are within the technology of ongoing and 
planned precision muon decay experiments. 

Finally, in a pioneering experiment\cite{16}, a dielectric 
suppression has been verified for Bremsstrahlung emission in scattering 
processes. This study gives us confidence in the reality of the 
material contributions to radiative corrections.

\end{document}